\documentclass[prl,aps,twocolumn,floats,nofootinbib,showpacs]{revtex4}
\usepackage{amsmath,amssymb,graphicx,psfrag}

\begin{document}
\renewcommand{\ni}{{\noindent}}
\newcommand{\dprime}{{\prime\prime}}
\newcommand{\be}{\begin{equation}}
\newcommand{\ee}{\end{equation}}
\newcommand{\bea}{\begin{eqnarray}} 
\newcommand{\eea}{\end{eqnarray}}
\newcommand{\la}{\langle}
\newcommand{\ra}{\rangle} 
\newcommand{\dg}{\dagger}
\newcommand\lbs{\left[}
\newcommand\rbs{\right]}
\newcommand\lbr{\left(}
\newcommand\rbr{\right)}
\newcommand\f{\frac}
\newcommand\e{\epsilon}
\newcommand\ua{\uparrow}
\newcommand\da{\downarrow}
\newcommand\rar{\rightarrow}
\newcommand\lar{\leftarrow}
\title{Drude weight in hard core Boson systems: possibility of a finite temperature ideal conductor}
\author{Gourab Majumder and Arti Garg} 
\affiliation{Condensed Matter Physics Division, Saha Institute of Nuclear Physics, 1/AF Bidhannagar, Kolkata 700 064, India} 
\vspace{0.2cm}
\begin{abstract}
\vspace{0.3cm}
We calculate Drude weight in the superfluid (SF) and the supersolid (SS) phases of hard core boson (HCB) model on a square lattice using stochastic series expansion (SSE). We demonstrate from our numerical calculations that the normal phase of HCBs in two dimensions can be  an {\it ideal conductor} with dissipationless transport. In two dimensions, when the ground state is a SF, the superfluid stiffness drops to zero with a Kosterlitz-Thouless  type transition at $T_{KT}$. The Drude weight, though is equal to the stiffness below $T_{KT}$, surprisingly stays finite even for temperatures above $T_{KT}$ indicating the non-dissipative transport in the normal state of this system. In contrast to this in a three dimensional SF phase, where the superfluid stiffness goes to zero continuously via a second order phase transition at $T_c$, Drude weight goes to zero at $T_c$, as expected. We also calculated the Drude weight in a 2-dimensional SS phase, where the charge density wave (CDW) order coexists with  superfluidity. For the SS phase we studied, superfluidity is lost via Kosterlitz-Thouless transition at $T_{KT}$ and the transition temperature for the CDW order is larger than $T_{KT}$. In striped SS phase where the CDW order breaks the rotational symmetry of the lattice, for $T > T_{KT}$, the system behaves like an ideal conductor along one of the lattice direction while along the other direction it behaves like an insulator. In contrast to this, in star-SS phase, Drude weight along both the lattice directions goes to zero along with the superfluid stiffness and for $T > T_{KT}$ we have a finite temperature phase of a CDW insulator.  
\vspace{0.1cm}
\end{abstract} 
\maketitle
\section{Introduction}
Superfluid phase of Bosons is a canonical quantum fluid just like the Fermi liquid phase of Fermions. 
One of the frontiers of quantum condensed matter physics is to explore quantum phases of Bosons in two-dimensions which are not superfluids. Lattice models of interacting bosons in two dimensions, such as Bose Hubbard model, which have been studied in past primarily as models for Josephson junction arrays~\cite{josephson} and in context of optical lattice experiments~\cite{optical lattice} and hard core bosons, which have been studied in context of the pseudogap phase of high $T_c$ superconductors~\cite{Assa}, are known to have insulating and superfluid phases~\cite{Troyer,Scalettar}. In some cases coexistence of charge density wave (CDW) order and superfluidity, which is known as the supersolid (SS) phase, is also seen~\cite{Scalettar,ss,Melko,chen,Mila,finiteT}. Most challenging phase, which is rarely seen, is gapless, compressible ``Bose-Metal" phase which breaks no symmetry whatsoever. There are very few examples of studies~\cite{Phillip,AP,Doniach,sheng} where ``Bose-Metal" phase has been realized.

Following Scalapino et. al.~\cite{Scalapino}, a superfluid (SF) phase is the one in which both the superfluid stiffness $\rho_s$ and the Drude weight $D$ are non-zero. In an insulating phase, e.g. in the CDW ordered phase, both $\rho_s=D=0$. Here $D$ is the delta function part of the charge conductivity $\sigma(\omega)=D\delta(\omega)+\sigma_{reg}(\omega)$ and  $\rho_s$ is given by the curvature of the thermodynamic limit of the free energy($\sim d^2{F}/d\phi^2$) with respect to a twist in boundary conditions($\phi$). Conventionally in a charged superfluid or a superconductor, $D$ remains non zero not only at zero temperature but at all temperatures below the transition temperature and is believed to be zero for temperatures above the transition temperature. In contrast to this, in a metal only at $T=0$ the Drude weight is defined to be non zero. With increase in temperature, the $\delta(\omega)$ peak in the conductivity gets broadened due to thermal fluctuations resulting in zero Drude weight. 
In this context, it is interesting to consider non-interacting Bose gas in one and two dimensions. In this case, at any finite temperature $\rho_s=0$ but since the current operator commutes with the Hamiltonian, the Drude weight remains finite at finite temperature. Therefore, non-interacting bosons in one and two dimensions at finite temperature are ``trivial" examples of {\it ideal conductors}~\cite{Sorella}. In this article, we explore the possibility of having a dissipatiionless ideal conductor of interacting bosons where $\rho_s$ goes to zero at certain transition temperature but $D$ remains non zero for a range of temperatures above the transition temperature.

To be specific, in this article we study the Drude weight in a model of hard core bosons (HCB), with nearest neighbour and next nearest neighbour hopping and repulsion terms, on 2d square lattice and cubic lattice. Phase diagram for this model of HCBs has been studied for a large range of parameters~\cite{Troyer,Scalettar,Melko,chen,finiteT}, but to the best of our knowledge, the Drude weight has not been calculated. We calculate the Drude weight in the SF, insulating and SS phase of this model using stochastic series expansion method~\cite{Sandvik_dloop,sandvik}. We demonstrate from our numerical calculations that the normal phase of HCBs in two dimensions can be an ideal conductor with dissipationless transport. Before going to the details of the paper, we summarize our main results below. 

 In two dimensional case, for the SF ground state, the superfluid stiffness drops to zero in the thermodynamic limit via Kosterlitz-Thouless~\cite{KT} type transition at $T_{KT}$~\cite{Troyer,Ding,Harada}. We found that 
 though the Drude weight is equal to the stiffness below $T_{KT}$ as expected, surprisingly it remains finite even for temperatures above $T_{KT}$ indicating the presence of an ideal Bose conductor with non dissipative transport in the normal phase of this two-dimensional SF. On the other hand, in three dimensions, where the superfluid transition is accompanied by the appearance of a true long-range order and the superfluid stiffness goes to zero via a continuous transition~\cite{Fisher1,Rigol} at $T_c$, we found that the Drude weight goes to zero at $T_c$. 

We also calculated the Drude weight in a SS phase on a 2d square lattice. We found that inspite of long range Ising order coexisting along with the superfluidity, $\rho_s$ drops to zero at $T_{KT}$ via a Kosterlitz-Thouless type transition. One of the SS phase we studied breaks the rotational symmetry and has striped CDW order only along one lattice direction (y direction). In this case, along y direction $D=\rho_s$ at all temperatures due to a spectral gap in the system and both $D$ and $\rho_s$ are zero above $T_{KT}$. On the other hand, along $x$ direction, $D$ remains finite even above $T_{KT}$. Therefore, the normal phase of the system is an ideal conductor along x direction while it is a finite temperature CDW ordered insulator along the $y$ direction. In the other SS phase we have studied, the CDW order survives in both the lattice directions. Therefore, along both the directions $D=\rho_s$ at all temperatures and both the quantities go to zero simultaneously at $T_{KT}$. 

The rest of this paper is organized as follows. In section I  we present the details of the model and the method used. Section II describes in detail the benchmarks on our code for calculation of the Drude weight showing comparison with earlier published results and with exact diagonalisation results on small system sizes. Section III describes the results in the CDW and the SF phases on a 2d square lattice followed up by our results for the SF phase on a 3d cubic lattice in section IV. In section V we present results for the SS phases on a 2d square lattice. We end this paper with conclusions and discussions on our work.     
\section{Model and Method}
 \vspace{-0.5cm}
We study hard core bosons on a square lattice described by the following  Hamiltonian 
\bea
H = -t\sum_{<ij>}(c_i^\dag c_j + c_j^\dag c_i)-t^\prime\sum_{<<ij>>}(c_i^\dag c_j + c_j^\dag c_i) \nonumber\\
+ V_1\sum_{<ij>} n_i n_j+ V_2\sum_{<<ij>>} n_i n_j
-\mu \sum_in_i
\label{hamil}
\eea
Here $t$ is the hopping amplitude from site i to its nearest neighbour site, $\mu$ is the chemical potential, $t^\prime$ is the next-nearest neighbour hopping amplitude and $V_1/V_2$, are the nearest neighbour and next neighbour repulsion terms, respectively. This model can be mapped onto the $S=\frac{1}{2}$ spin model using the exact mapping $ S_i^{\dag}= a_i^{\dag}$ and $S_i^{z} = n_i-1/2$.  
In the spin language one gets the extended XXZ model
\bea
H = -t\sum_{<ij>}(S_i^\dag S_j^{-} + S_j^\dag S_i^{-})-t'\sum_{<<ij>>}(S_i^\dag S_j^{-} + S_j^\dag S_i^{-})\nonumber\\ 
+ V_1\sum_{<ij>} S_i^z S_j^z + V_2\sum_{<<ij>>} S_i^z S_j^z 
- h \sum_{i}S_i^z
\eea
where $h=\mu-2V_1-2V_2$.
This model has been studied earlier extensively using stochastic series expansion.  
With only nearest neighbour terms, this model is known to have a SF phase and  an insulating phase with a CDW order. The quantum phase transition from SF to CDW phase can be attained either by tuning the repulsion term or the chemical potential~\cite{Troyer,Scalettar}. Finite temperature phase diagram for this model has also been studied~\cite{Troyer}. Upon increasing temperature the superfluid stiffness drops to zero with a Kosterlitz-Thouless (KT) type transition at $T_{KT}$, just like in the model with only nearest neighbour hopping for the hard core bosons~\cite{Ding,Harada}. On the other hand, in three dimensions, the superfluid transition is accompanied by the appearance of true long-range order and the superfluid stiffness goes to zero with a continuous transition~\cite{Fisher1,Rigol}. The full model with next neighbour interactions is known to have exotic supersolid phases~\cite{Melko,chen,Mila,finiteT}. 
In this paper, we study transport properties, mainly the Drude weight in the charge conductivity, of all these phases at finite temperature using SSE with directed loop update \cite{Sandvik_dloop}. Below we describe how the Drude weight and superfluid stiffness can be calculated within linear response theory (Kubo formula) using SSE. 

\subsection{Drude weight and superfluid stiffness}
Superfluid density ($\rho_s$) is given by the curvature of the thermodynamic limit of the free energy($\pi /N d^2{F}/d\phi^2$)
with respect to a twist in boundary conditions($\phi$). To evaluate it within SSE, we use the kubo formula representation of this quantity~\cite{Scalapino}
\bea
\rho_s= 
         \la -K_x\ra - Re\Lambda(q_x = 0 , q_y\rightarrow 0 , i\omega_m = 0)\nonumber\\
         =\frac{\la W^2\ra}{\beta}
\label{rhos}
\eea  
       
Here $\la -K_x \ra$ is the kinetic energy, $\Lambda(\vec{q},i\omega_m)$ is the current current correlation function and $\la W^2\ra$ is the winding number. 
For the model with nearest and next near neighbour hopping, $K_x$ and the current operator $J_x$ are given by 
\bea
 K_x= -t\sum_{i}(c_i^{\dag}c_{i+\hat{x}}+c_{i+\hat{x}}^{\dag}c_i)\nonumber\\
-t^\prime\sum_{i}(c_i^{\dag}c_{i+\hat{x}\pm\hat{y}}+c_{i+\hat{x}\pm\hat{y}}^{\dag}c_i)]  \\
J_x(q=0) = it\sum_{i}(c_i^{\dag}c_{i+\hat{x}}-c_{i+\hat{x}}^{\dag}c_i)\nonumber\\
+it^\prime\sum_{i}(c_i^{\dag}c_{i+\hat{x}\pm\hat{y}}-c_{i+\hat{x}\pm\hat{y}}^{\dag}c_i)
\label{kx_jx}
\eea
Here $\hat{x}$ and $\hat{y}$ denote unity vectors along the X and Y axis of the lattice, respectively.

The Drude weight $D$ is obtained by taking the transport limit of the Kubo formula~\cite{Scalapino,Mahan}, namely, 
\bea
D= 
         \la -K_x\ra - \Lambda(q_x = 0 , q_y = 0 , \omega \rar 0) 
\label{D}
\eea

The current-current response function is
          \bea 
          \Lambda_{xx}(\vec{q},i\omega_{m}) = \int_{0}^{\beta}d\tau e^{i\omega_m\tau}
          \la J{_{x}}(\vec{q},\tau)J{_{x}}(-\vec{q},0)\ra
	   \label{lambda}
          \eea 
          $\omega_m$ is the Matsubara frequency given by $2\pi m/\beta$ where $m$ is any integer and $\beta$ the inverse of temperature. 

We follow work by~\cite{Heidarian} to evaluate this expression within SSE. Let us use the symbol $H_b^{+}$ for $c_i^\dag c_{i+\hat{x}}$. Then calculation of  $\Lambda$ includes the imaginary-time($\tau$) ordered average of the product $\la H_{b_1}^{\sigma_1}(\tau)H_{b_2}^{\sigma_2}(\tau=0)\ra$  where $b_1$ and $b_2$ are bond indices and $\sigma_{1,2}=\pm$.  Within SSE, time ordered average of any two such local operators \cite{sandvik} can be represented as follows
      \bea &&\!\!\!\!\!\!\!\!\la H{_{b_2}^{\sigma_2}}(\tau)H{_{b_1}^{\sigma_1}}(0)\ra \nonumber \\
        &&\!\!\!\!\!\!\!\!=\frac{\sum_{k}\sum_{n,m}\frac{(\tau-\beta)^n(-\tau^m)}{n!m!} 
         <\Psi_k|H^nH{_{b_2}^{\sigma_2}}H^mH{_{b_1}^{\sigma_1}}|\Psi_k>_W}{Z}  \nonumber \\
        &&\!\!\!\!\!\!\!\!\!\! = \la \sum_{m=0}^{n_s-2}\frac{(\beta-\tau)^n(\tau^m)}{\beta^n}\frac{(n-1)!}{(n-m-2)!m!}N^{\sigma_1\sigma_2}_{b_1b_2}(m)\ra_W \label{drudsse}
\eea
         
where we have the summations over n and m coming out from the Taylor expansion of $e^{(-\beta+\tau)H}$ and $e^{-\beta H}$ and $N^{\sigma_1\sigma_2}_{b_1,b_2}(m)$ is the number of times such a combination
with m non-identity operators in between, appears in the operator sequence in SSE. Fourier transform of Eq.~\ref{drudsse} from $\tau$ to matsubara frequency $\omega_m$ space yields:
    \bea
      \frac{-1}{\beta}\sum_{\sigma_1,\sigma_2=\pm}\sigma_1\sigma_2\sum_{m} F{_1^1}(m+1,n_s;2i\pi n)N_{b_1b_2}^{\sigma_1\sigma_2}(m)
      \eea
 where $F{_1^1}$ is the confluent hypergeometric function of first order~\cite{Heidarian}.\\
 \begin{figure}[t!]
      \begin{center}      
     \includegraphics[width=7.0cm,height=7.0cm,angle=0]{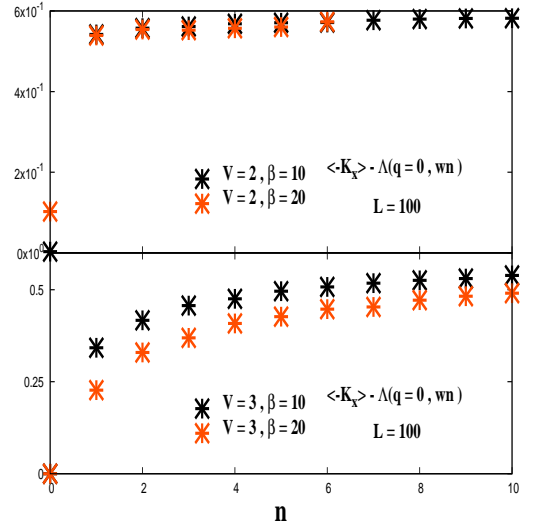}
     \caption{Results for 1D $t-V_1$ model to check our SSE code for the calculation of the Drude weight. Results are shown for two values of the
              nearest neighbour repulsion, namely, $V_1 = 2t$ and $3t$. $V_{1c} = 2t$ is the critical parameter below which $D$ is non-zero. For $V_1 = 3t$, $\la -K_x\ra -lim_{\omega_n\rightarrow 0}\Lambda_{xx}(i\omega_n) \rightarrow 0$.}
     \label{1d}
      \end{center}
      \end{figure}

After calculating the current current correlation function $\Lambda(q,i\omega_n)$ this way within SSE, one calculates the Drude weight using Eq.~\ref{D}.  
The analytical continuation of the current-current correlation function $\Lambda$, given in Eq.~\ref{lambda}, is valid in the continuous upper complex plane, including the imaginary axis at frequencies different from Matsubara frequencies. One can therefore take the limit $\omega \rightarrow 0$ for $\Lambda$ either along the real axis, or purely on the imaginary axis, even at finite temperature. Here, we extrapolate the imaginary axis data to obtain the Drude weight without carrying out analytic continuation. This method has earlier been used extensively for calculation of Drude weights in 1-d systems~\cite{Heidarian,1dhubbard}. In order to carry out the extrapolation to $\omega_n \rightarrow 0$ we fit real part of $\Lambda(i\omega_n)$ vs $n$ with polynomial and Lorentzian functions. The reason of choosing the Lorentzian is that the current-current correlation function $\Lambda(i\omega_n)$ is a well-behaved function on the imaginary axis being the sum of Lorentz curves:
\be
\Lambda(i\omega_n)=\sum_j c_j \frac{\Delta_j}{\omega_n^2+\Delta_j^2}
\ee
We approximate it by a finite series 
\be
\Lambda(i\omega_n) = \frac{a}{\omega_n^2+b^2}+\frac{c}{\omega_n^2+d^2}
\ee
and determine the constants $a,b,c,d$. This method, as explained in ~\cite{1dhubbard}, is a well known method for extrapolation of current current correlation functions and has been extensively used for determining Drude weight at finite T for 1D systems. In many cases a single Lorentzian $a/(b^2+\omega_n^2)$ provides a good fit to the data. Details of comparison of polynomial and Lorentzian fits for various data sets are shown in Appendix A. 
\section{Benchmarking the code}
We cross checked our SSE results for the Drude weight with earlier work on 1-dimensional $t-V_1$ model~\cite{Heidarian}. For this case, there is a critical point
($V_{1c}/t = 2$) below which the system is a SF at $T=0$. For $V_1 > 2t$, the system is an insulator with a CDW order. Though $\rho_s$ is zero at any finite temperature in this 1D system, ~\cite{Heidarian} showed that the Drude weight is finite even at finite temperatures for $V_1 \le 2t$. In the CDW phase, for $V_1 > 2t$, the Drude weight is zero in the thermodynamic limit at any temperature. Fig.~\ref{1d} shows one such result where the current-current correlations (in addition with the kinetic energy term) against Matsubara frequency are shown for two values of $V_1$. For $V_1 = 2$ the $\la -K_x\ra -\Lambda(q=0,\omega_n = 0)$,  which gives the superfluid density ($\rho_s$), is zero for $T=0.1t$ (though for $\beta=20$, for $L=100$, $\rho_s$ is still non zero but will tend to zero upon increasing the system size), whereas, the values for non-zero $\omega_n$ are finite. The extrapolation of $\la -K_x\ra -\lambda(i\omega_n)$ to $\omega_n \rightarrow 0$ gives a non-zero Drude weight for this case.  On the other hand, for $V_1 = 3$, as shown in the bottom panel of Fig.~\ref{1d}, the extrapolated value of $\la -K_x \ra -\Lambda(q=0,\omega_n)$ goes to zero and also matches with the value at $\omega_n=0$, implying a zero value for the Drude weight and the stiffness. These results are consistent with published results in ~\cite{Heidarian} and provide a test to our Drude weight code at low temperature. 
\begin{figure}[h!]
\begin{center}
\hspace{-1cm}
\includegraphics[width=9.0cm,angle=0]{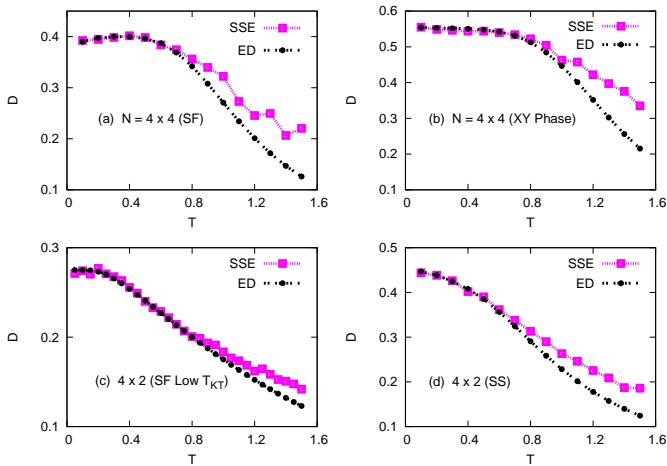}
\caption{Comparison of Drude weight calculated within SSE and ED for small system size in different phases. Panel [a] shows the $D$ vs $T$ plots for the results obtained on a $4\times4$ lattice for the SF phase in XXZ model. Panel [b] shows the results for the SF in the XY model on a $4\times4$ lattice. Panel [c] and [d] shows comparison of SSE and ED results on a $4\times2$ lattice for the low $T_{KT}$ SF phase and the SS-I phase respectively. Please note that in all the phases, $D_{SSE}-D_{ED} \le 0.01$ for $T \le 0.8-1$ in units of $t$.}
\label{ed}
\end{center}
\end{figure}
\\
Getting reliable Drude weight at higher temperatures (due to the increase in the minimum value of $\omega_m$) is difficult using this method. 
 In order to have an idea about the maximum range of temperatures up to which the extrapolation works, we cross-checked our SSE data against Exact Diagonalization (ED) results for small system sizes. Results from this comparison are shown in Fig.~\ref{ed}. By calculating the Kubo formula exactly for a small system size in ED where no extrapolation is required to obtain the Drude weight, we could estimate errors in the corresponding SSE calculation of the Drude weight which requires extrapolation in Matsubara frequency.
In exact diagonalisation, using the eigenvalues and the eigenvectors, one can calculate the Drude weight from the Lehmann representation of the Kubo formula and one arrives at the following expressions~\cite{Mahan,1dhubbard}:
\bea
D(T) = -<K_x> - \frac{2}{L}\sum_{n,m}^{E_n\ne E_m}\frac{p_n}{E_m-E_n}|<n|J_x(0)|m>|^2 \nonumber \\
\eea
Here $|n\ra $ is the eigenvector of the Hamiltonian with eigenvalue $E_n$, i.e., $H|n\ra =E_n|n\ra$ and $p_n=\exp(-\beta E_n)/Z$ with $Z$ being the partition function. Superfluid density can be calculated as
\bea
\rho_s(T) = -<K_x> - \frac{2}{L}\sum_{n,m}^{E_n\ne E_m}\frac{p_n}{E_m-E_n}|<n|J_x(0)|m>|^2 \nonumber\\
\hspace{-1cm}-\frac{\beta}{L}\sum_{n,m}^{E_n = E_m}P_n|<n|J_x(0)|m>|^2 ~~~~~~~~~~~~~~~~(11)\nonumber 
\eea

As shown in Fig.~\ref{ed}, in all the phases, for $T/t \le 0.8-1.0$, $D$ within SSE and ED calculation matches very well.  
In the following section we describe our results for the CDW and SF phases in two dimensions.  
 \begin{figure}[h!]
 \begin{center}      
 \includegraphics[width=6.0cm,height=6.0cm,angle=0]{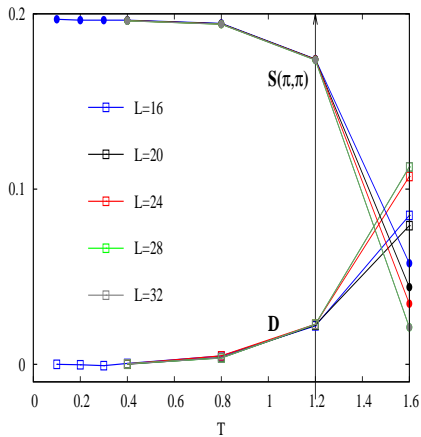}
 \caption{Structure factor $S(\pi,\pi)$ and the Drude weight $D$ vs $T$, for various system sizes, in the CDW phase. In an insulator, $D$ should be zero at all temperatures. From our SSE calculation, $D \le 0.01$ for $T\le1.0t$. This gives an idea about the maximum range of temperature up  to which our SSE results for $D$ are reliable.}
 \label{cdw}
 \end{center}
 \end{figure}

\section{Results in 2d} In this section we describe our results for various phases seen in the HCB model in Eq.~\ref{hamil}. In order to have an idea about the maximum range of temperature up  to which our Drude weight calculation is reliable, we first present our results for the CDW phase followed up by details of other phases.
\begin{figure}[h!]
\begin{center}
\includegraphics[width=7.5cm,angle=0]{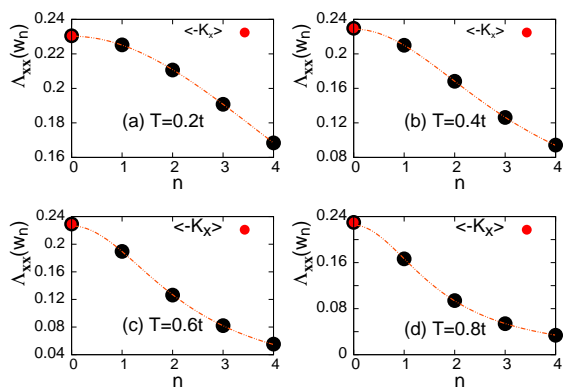}
\caption{Extrapolation plots for $\Lambda_{xx}(i\omega_n)$ vs $n$ at various temperatures in the CDW phase. Note that the extrapolated value of $\Lambda(i\omega_n)$ is equal to its value at $\omega_n=0$, implying that $\rho_s=D$. Further $\Lambda(i\omega_n=0) =\la -K_x\ra$, which means both $D$ and $\rho_s$ are zero in the CDW phase.}
\label{cdw2}
\end{center}
\end{figure}
 
\subsection{Drude weight in the CDW Phase} 
A staggered charge order appears at half filling in the ground state of model in Eq.~\ref{hamil} for $V_1=3t$ and $h=0$ with no next nearest neighbour hopping and repulsion~\cite{Troyer}. In terms of the spin model this CDW phase is equivalent to the Antiferromagnetic phase. In terms of bosons this phase is an insulator having a gap in the single particle excitation spectrum. Therefore, both the superfluid stiffness and the Drude weight must be zero at all temperatures ~\cite{Scalapino}. 

 Fig.~\ref{cdw} shows the structure factor $S(\pi,\pi) = \sum_{i,j} (-1)^{i+j}\la S_z(i)S_z(j)\ra$, which represents the staggered checkerboard charge order in this system, and the Drude weight $D$ vs $T$ for various system sizes. The CDW order parameter reduces with increase in temperature and goes to zero continuously at a transition temperature of $T_c=1.5t=0.5V_1$~\cite{Troyer}. 
The Drude weight is indeed zero ($D\le 0.01$) up  to $T \le 1.0t$ within our SSE calculations. Hence, we can say that our results are up to expectations for temperatures below $1.0t$, which is also consistent with benchmarking of our SSE data against ED for small system sizes. Detailed plots of the current-current correlation function $\Lambda(i\omega_n)$ vs $\omega_n$ are shown in Fig.~\ref{cdw2}, along with the kinetic energy values for various temperatures and system sizes. Notice that at all temperatures, $lim_{\omega_n \rightarrow 0}\Lambda(i\omega_n)= \Lambda(\omega_n=0)$, which implies that $D=\rho_s$. Also the extrapolated value of $\Lambda(i\omega_n)$ is equal to $\la -K_x\ra$, implying that both $D$ and $\rho_s$ are zero in the insulating CDW phase.  
\begin{figure}[h!]
\begin{center}
\includegraphics[width=6.0cm,height=6.0cm,angle=0]{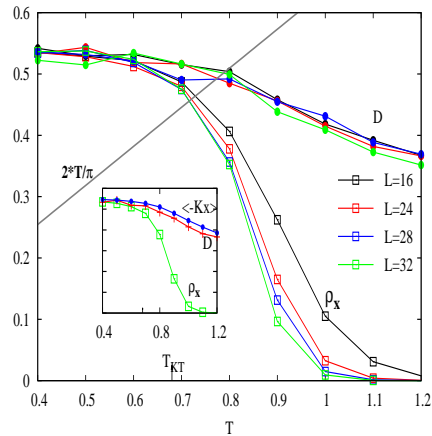}
\caption{$\rho_s$ and $D$ vs $T$ for the quantum XY model. Drude weight $D$ remains non-zero above $T_{KT}$ and shows almost no change with the system size. Inset shows the kinetic energy $\la-K_x\ra$, Drude weight($D$) and superfluid density ($\rho_s$) vs $T$ for $L=32$.}
\label{xy}
\end{center}
\end{figure}

\subsection{Superfluid phase} 
The generic model in Eq.~\ref{hamil} shows a superfluid ground state for a wide range of parameters~\cite{Troyer,Scalettar,Melko,chen}. The SF phase survives at finite temperature up  to $T_{KT}$   where $\rho_s$ goes to zero with a universal drop, of Kosterlitz-Thouless type transition~\cite{KT}, in the thermodynamic limit.  For $T \ll T_{KT}$, where the linear spin wave approximation holds good, $\rho_s=D = \la -K_x\ra$, because the current-current correlation function is zero. But how the Drude weight behaves at higher temperatures is not known. Our numerical calculation, results from which are presented in detail below, shows that in all the SF phases, $D=\rho_s$ for $T < T_{KT}$ and for $T> T_{KT}$, $D$ starts deviating from $\rho_s$. For $T \ge T_{KT}$, though $\rho_s \rightarrow 0$ in the thermodynamic limit, $D$ stays finite even in the thermodynamic limit for a large range of temperature beyond $T_{KT}$, which implies that the normal phase in this 2 dimensional system has dissipationless transport.
\begin{figure}[h!]
\begin{center}
\hspace{-0.4cm}
\includegraphics[width=8.0cm,angle=0]{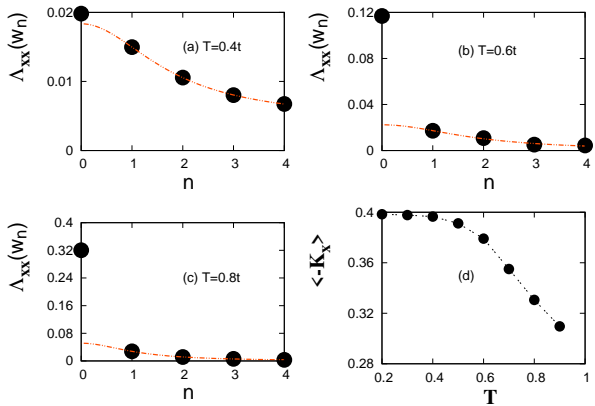}
\caption{Extrapolation plots for $\Lambda_{xx}(i\omega_n)$ vs $n$ at various temperatures for XY model. Note that the extrapolated value of $\Lambda_{xx}(i\omega_n)$ is equal to its value at $\omega_n=0$ at low temperatures, which implies that $\rho_s=D\sim \la -K_x\ra$ at low $T$. But for higher temperature values, $lim_{n\rightarrow 0}\Lambda_{xx}(i\omega_n) \ne \Lambda_{xx}(\omega_n=0) \ne \la -K_x\ra$, which means that $D \ne \rho_s$ and both of these quantities are different from $\la -K_x\ra$.}
\label{xy2}
\end{center}
\end{figure}
 Since our extrapolation method of evaluating Drude weights becomes erroneous for higher temperatures, in order to check our observation about non zero Drude weight above $T_{KT}$, we analyze SF phase not only in the pure XY model ($T_{KT}=0.68t$) but we also looked for the SF phases  with lower $T_{KT}$. Below we present in detail the results for all the SF phases we have studied.     
\begin{figure}[h!]
\begin{center}
\includegraphics[width=6cm,height=6cm,angle=0]{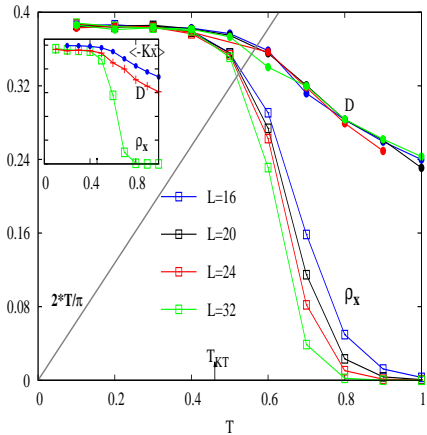}
\caption{$\rho_s$ and $D$ vs $T$ for the SF phase of the quantum XXZ model. Results are shown for 2d square lattice of various sizes, namely, $L=16,20,24$ and $32$. Inset shows the kinetic energy $\la-K_x\ra$, Drude weight($D$) and the superfluid density ($\rho_s$) vs $T$ for $L=32$.}
\label{xxz}
\end{center}
\end{figure}
\subsubsection {XY model}
First we study the simplest model with only the nearest neighbour hopping term for hard core bosons. All other couplings in Eq.~\ref{hamil} are set to zero in this case. In the spin language, this maps to the pure quantum XY model which has been rigorously studied using SSE~\cite{Troyer,Scalettar,Rigol} and is known to have a Kosterlitz-Thouless type transition at $T_{KT}=0.68t$~\cite{Ding,Harada}. Fig.~\ref{xy} shows the plot of the superfluid density $\rho_s$, Drude weight $D$ and the kinetic energy $\la -K_x\ra$ vs temperature ($T$) for various system sizes. We see that for $T < T_{KT}$, $\rho_s \sim D$ both being bounded from above by the $\la -K_x\ra$. For $T > T_{KT}$, though $\rho_s$ goes to zero in the thermodynamic limit, $D$ shows a much slower decrease with $T$. Further, $D$ does not show any significant system size dependence and remains non zero even in the thermodynamic limit, which implies that the normal phase of this system is an ideal conductor.

 Detailed plots for the current current correlation function $\Lambda(i\omega_n)$ vs $\omega_n$ are shown in Fig.~\ref{xy2}. For $T \ll T_{KT}$, deep in the SF phase, $lim_{\omega_n\rightarrow 0}\Lambda(i\omega_n) \sim \Lambda(\omega_n=0)\ll \la -K_x\ra$, and thus $D\sim \rho_s$, both being non-zero. As $T$ increases, still being in the SF phase, $lim_{\omega_n\rightarrow 0}\Lambda(i\omega_n) < \Lambda(\omega_n=0)$, making $D > \rho_s$. Same trend for $D$ continues for $T > T_{KT}$ where $\Lambda(\omega_n=0)\rightarrow \la -K_x\ra$ making $\rho_s\rightarrow 0$ in the thermodynamic limit.   
Note that $T_{KT}=0.68t$ for this phase which is very close to the $T_{max}=0.8t$ within which we can get reliable Drude weight. Below we present our results for the SF phases with lower values of $T_{KT}$.
\begin{figure}[h!]
\begin{center}      
\includegraphics[width=6cm,height=6cm,angle=0]{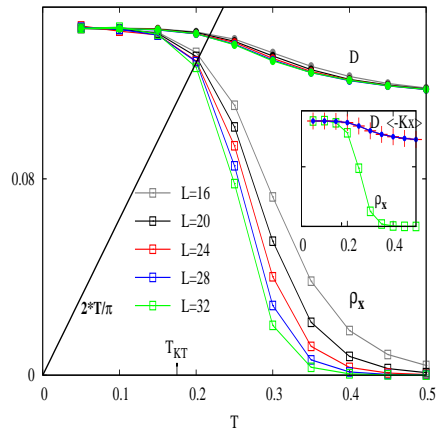}   
\caption{$\rho_s$ and $D$ vs $T$ evaluated for the low $T_{KT}$ SF phase ($t=0.9$,$t'=0.1$,$V_1=1,V_2=4.5,h=14.0$). Results are shown for 2d square lattice of various lengths, namely, $L=16,20,24,28$ and $32$. Inset shows the kinetic energy $\la-K_x\ra$, Drude weight($D$) and the superfluid density ($\rho_s$) vs $T$ for $L=32$.}
\label{lowKT}
\end{center}
\end{figure}
   
\subsubsection{XXZ model}
We study another SF phase, which is the ground state of the XXZ model with $t = 1 , V_1 = 3 , h = 6 $. Here the system shows a KT type transition at $T_{KT}=0.47t$ which was concluded from the conventional logarithmic scaling behavior~\cite{Troyer} of the transition temperature.
In Fig.~\ref{xxz}, we show our finite temperature results for $\rho_s$ and $D$ for this phase. It can be clearly seen that at temperatures much higher than $T_{KT}$ Drude weight survives and shows a slow decreasing behaviour with $T$ much like the kinetic energy. Also the system size dependence for $D$ is much weaker compared to that of $\rho_s$ implying a non zero $D$ even in the thermodynamic limit for $T > T_{KT}$.  
\subsubsection{Superfluid phase with much lower $T_{KT}$}
We extended our analysis for another superfluid phase, having a much lower transition temperature. To do that we turned on $t'$ and $V_2$ and choose their values according to the existing literature~\cite{Melko} to be  $t=0.9,t'=0.1,V_1=1.0,V_2=4.5,h=14.0$ where the system exhibits a superfluid ground state. 
A closely related phase is studied in ~\cite{finiteT}. 
As shown in Fig.~\ref{lowKT}, the superfluidity is lost with a Kosterlitz-Thouless type transition at $T_{KT}=0.17t$. This phase is realized at a particle density of $n=0.93$ which means holes, the carriers of super fluidity, have very low density $0.07$ here. At this low density, neither the hard core constraint and nor the effect of nearest or next nearest neighbour repulsion is significant and effectively we have a gas of non interacting bosons in 2d. For an ideal Bose gas in 2d $\rho_s$ is zero at any finite $T$ while the Drude weight $D$ is non zero being equal to $\la -K_x\ra$~\cite{Sorella}. Note that here we are in close proximity of the ideal  two dimensional Bose gas, which is indicated by low value of $T_{KT}$ and the Drude weight data shown in Fig.~\ref{lowKT}. As shown in the inset, for all $T$ studied, $D \sim \la -K_x\ra$ as expected for this low density phase. Just like in the high $T_{KT}$ superfluid phases, Drude weight is equal to $\rho_s$ for $T < T_{KT}$ but stays non zero for $T > T_{KT}$ without showing any significant system size dependence. Note that since $T_{KT}$ for this system is much smaller than the maximum $T$ limit within which our extrapolation errors are under control, our analysis for $D$ above $T_{KT}$ (up  to $T=0.5t$) is very reliable and supports our proposal of the normal phase being an ideal conductor.

After detailed demonstration of results for the 2d SF phase, we come to the question why the Drude weight remains non zero even above $T_{KT}$ where the superfluid stiffness drops to zero in the thermodynamic limit? We propose the following explanation for this observation. There are basically two types of excitations possible possible in this system, namely, the spin wave excitations and the vortex excitations. This is well known for the corresponding classical model~\cite{KT} and vortex excitations have also been observed in hard core bosons ~\cite{vortex}. At very low temperature, spin wave excitations are present while the vortex-anti vortex pairs are bound, having effectively no vortex excitations. In this regime, $D=\rho_s=\la -K_x\ra$. As $T$ increases, more spin waves are excited and start interacting with each other. For $T\ge T_{KT}$, due to unbound vortices $\rho_s$ drops to zero. But somehow $D$ is not suppressed by the presence of vortices. One reason for it might be that the vortices near $T_{KT}$ are ballistic. This hypothesis is made in original paper by Kosterlitz and Thouless~\cite{KT}. Another reason might be that the Drude weight, which is obtained from the long wavelength limit, before taking $\omega\rightarrow 0$ limit, of the Kubo formula, does not feel the presence of vortices which are local excitations though it might be affected by interaction between spin waves and vortices. This picture can be confirmed by studying the SF phase of HCB's in three dimensions, where the system has true long range order and it undergoes a continuous transition (instead of the KT transition) from the SF phase to the normal phase, having spin waves as the only relevant excitations. With this motivation, we study the quantum XY model on a cubic lattice in the following section and compare results with the 2d case.

\begin{figure}[h!]
\begin{center}      
\includegraphics[width=7cm,angle=0]{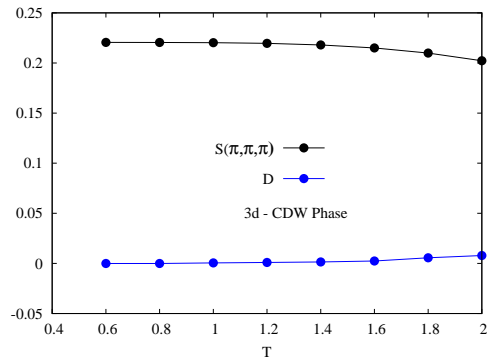}     
\caption{Temperature dependence of $S(\pi,\pi,\pi)$ and $D$ for the CDW ordered phase in three dimensions for $L=10$. $ D \le 0.01$ for $T< 2.2t$.}
\label{3d-cdw}
\end{center}
\end{figure}

\section{Results in three dimensions}
Before presenting our results for the SF phase, we first study the CDW ordered phase in 3d realized for $t=1.0,V_1=3.0$ in Eq.~\ref{hamil} keeping all other couplings to be zero. This will help us in estimating the error bars in calculation of $D$ and also in finding the maximum temperature $T$ up  to which calculation of Drude weight is reliable for this system. Fig.~\ref{3d-cdw} shows the structure factor $S(\pi,\pi,\pi)$ and the Drude weight $D$ vs $T$. As seen clearly that $D \le 0.01$ for $T \le 2.2t$. Note that in terms of the bandwidth $W$, which is $4t$ for the square lattice and $6t$ for the cubic lattice, the range of $T$ up  to which we get reliable results for $D$ is roughly the same in two and three dimensions. 
\begin{figure}[h!]
     \begin{center}      
     \includegraphics[width=8cm,angle=0]{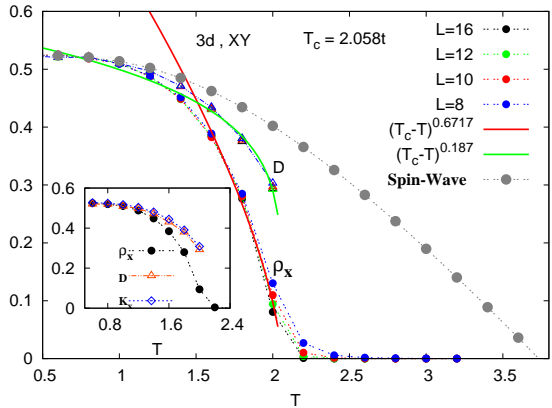}     
     \caption{Temperature dependence of  $\rho_s$ and $D$ for  a quantum XY model in three dimensions. $\rho_s$ satisfies the scaling form $(T_c-T)^\nu$ with $\nu=0.67$ and $T_c\sim 2.058t\pm0.01t$. The Drude weight $D$ is extrapolated to higher temperatures using the scaling form $D \sim (T_c-T)^{\nu(1-z)}$ with $\nu(1-z)\sim 0.187$. Note that $D$ is zero for temperatures larger than $T_c$. Linear spin wave results, within which $\rho_s=D=\la -K_x\ra$, are shown for comparison.}
     \label{3d-T}
     \end{center}
     \end{figure} 
 
Now we discuss our results for the quantum XY model in 3d. Temperature dependence of $\rho_s$ and $D$ for various cubes of length $L$, obtained from SSE is shown in Fig.~\ref{3d-T}. Note that phase transition in 3D XY model is in 3 d O(2) class where the superfluid stiffness goes to zero via a continuous transition at $T_c$.  For this universality class, the superfluid stiffness near the transition temperature for $T < T_c$ behaves as $\rho_s \sim (T_c-T)^{(d-2)\nu}$ with $\nu=0.6717$~\cite{Fisher1,Rigol}. From the scaling of our SSE data in Fig.~\ref{3d-T}, we found $T_c=2.058t\pm0.01t$ which is close to the value reported earlier~\cite{Rigol}. For very low temperature, $D=\rho_s=\la -K_x\ra$. This is the regime where linear spin wave theory works well. As $T$ increases, deviation from linear spin wave theory occurs due to enhanced interaction between spin waves, which is not incorporated in linear spin wave theory. As a result, $D$ starts deviating from $\rho_s$. In order to see whether $D$ goes to zero at $T_c$ or not, we derive below a scaling form for $D$ and then check it against our SSE data. 

As defined earlier,the charge conductivity $\sigma(\omega)=\frac{i}{\omega+i\eta}D+\frac{\Lambda^{\prime\prime}}{\omega}$. For $T> T_c$, $D$ is zero while $\Lambda^{\prime\prime}$ remains non zero, so we consider scaling of first term in the expression for $\sigma(\omega)$ and we label it as $\sigma_1(\omega)$. With an extension of expression in~\cite{Fisher}, we get 
\be
\sigma_1(\omega)\sim \xi^{z+2-d} f(\omega \tau_\sigma)
\ee
 Here the difference compared to the expression in Fisher et. al paper~\cite{Fisher} is that we have considered two time scales to write this expression. One is the correlation time $\tau\sim \xi^z$ with $z$ being the dynamical exponent. Another is the transport time $\tau_\sigma$ scale which decides the rate of scattering between the particles.
In general, $\tau_\sigma$ must be different from the correlation time scale $\tau$. We assume that $\tau_\sigma \sim \xi^{z_\sigma}$ where $z_\sigma$ in general will be different from $z$. Further, following~\cite{Fisher}, the function $f(x)=constant$ for $T>T_c$ implying zero Drude weight for $T> T_c$ and for $T< T_c$, $f(x)=i/x$ giving $\delta(\omega)$ term in real part of conductivity. Thus the Drude weight goes as $\xi^{z+2-d-z_\sigma}$ which implies $D \sim (T_c-T)^{\nu(d-2-(z-z_\sigma))}$. Note that only for $z=z_\sigma$, $D$ will follow the same scaling form as $\rho_s$. We did fitting of the Drude weight data and found $\nu(1-(z-z_\sigma))\sim 0.187$ fits the data quite well which gives $z-z_\sigma \sim 0.72$. There is no consensus on values of $z$ for 3d XY model. Values in range $0.07-1.6$ have been reported in various numerical calculations~\cite{scalingXY}. 

Interestingly, since the SSE data for the Drude weight fits very well to the scaling form for 3d O(2) class, we see that $D$ goes to zero at $T_c$ along with $\rho_s$ which is in clear contrast to the 2d case where $D$ remains non zero for a large range of $T$ above $T_{KT}$. This numerical observation is consistent with our intuitive picture that the Drude weight $D$ is governed primarily by the spin wave excitations and not by the vortex excitations.  

Before closing the section on results, below we present our results for an interesting, exotic phase, namely the Supersolid phase.
\begin{figure}
\begin{center}
\includegraphics[width=8cm, angle=0]{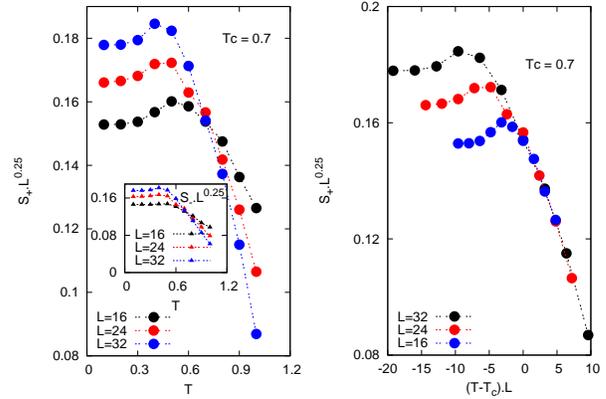}
\caption{Structure factors for the CDW order in the SS-I phase. Left panel shows $S_{+}$ vs $T$ for various system sizes while the right panels shows the scaling behaviour indicating $T_c\sim=0.7t$. Note that both $S_{+}$ and $S_{-}$ are non zero, as shown in the inset, and of the same strength at all temperatures. This implies that the system has the CDW order only along one of the lattice direction, either x or y.}
\label{ss1_S}
\end{center}
\end{figure} 
\section{Supersolid Phase in 2d}
Finally, we turn our attention towards a more complicated phase, namely, the supersolid phase defined to be a homogeneous mixture of both the superfluid phase and the CDW phase. In this section we will present results for two supersolid phases we have studied. 
\begin{figure}
\begin{center}
\includegraphics[width=8cm, angle=0]{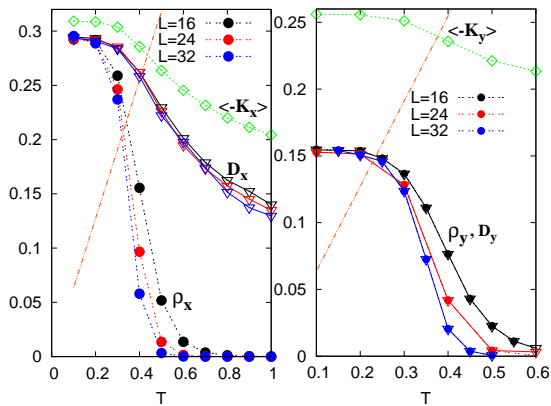}
\caption{Left panel shows the superfluid stiffness and the Drude weight vs $T$ along x direction for various system sizes. Note that the behaviour is very similar to that of a 2d SF phase where $\rho_x$ drops to zero at $T_{KT}$ while $D_x$ remains non zero for $T > T_{KT}$. Along this direction $\rho_x =D_x\sim\la -K_x\ra$ at low $T$. The right panel shows the $\rho_y$ and $D_y$ along the y direction vs $T$. At very low $T$ itself, $\rho_y= D_y \ne \la -K_y\ra$ due to non zero CDW order along this direction. Note that $\rho_y=D_y$ at all temperatures and $D_y$ goes to zero at $T_{KT}$ along with $\rho_y$.}
\label{ss1_T}
\end{center}
\end{figure} 

\subsection{Supersolid-1} 
We choose parameter points to be $t = 0.9, t' = 0.1, V_1 = 4.5, V_2 = 4.5, h = 9.0$ where the average density for bosons is $2/3$ and a striped SS phase has been reported~\cite{Melko}. The finite temperature phase diagram for this system has not been studied earlier though for a very closely related in parameter space and qualitatively similar striped SS phase it has been studied~\cite{finiteT} at finite temperature.  We first study the structure factors corresponding to various charge orderings in this system. We calculated the structure factor $S(Q)=\sum_{i,j} \exp(i Q\cdot (r_i-r_j)) \la S_z(i)S_z(j)\ra$ with $Q$ along the symmetry directions, namely, $(\pi,0), (0,\pi)$ and $(\pi,\pi)$.  
Within SSE it is not possible to calculate $S(\pi,0)$ or $S(0,\pi)$ separately, but one calculates $S_{+}=S(\pi,0)+ S(0,\pi)$ and  $S_{-}=|S(\pi,0)-S(0,\pi)|$. We found that in the ground state the CDW order breaks the rotational symmetry of the lattice. There is no order along the $(\pi,\pi)$ ordering wavevector and $S_{+} \sim S_{-}$ which implies that only one of $S(\pi,0)$ or $S(0,\pi)$ is non zero and other one is vanishingly small. With increase in temperature $T$, both $S_{\pm}$ are more or less constant for low $T$ and then starts decreasing with the two curves remaining parallel to each other as shown in the left panels of Fig.~\ref{ss1_S}. We noticed a slight increase in $S_+$ before it starts decreasing with $T$, which might be due to small competing order like $S(\pi,\pi) \ll S_{\pm}$. To get an estimate about $T_c$, we did scaling (shown in the right panel of Fig.~\ref{ss1_S}) of $S_{\pm}$ assuming that the system belongs to the Ising class ~\cite{finiteT} and estimated that the transition temperature for the charge order is $T_c\sim 0.7t$ for both the $x,y$ components.  

This system also has a non zero superfluid stiffness as shown in Fig.~\ref{ss1_T}. Since the charge order breaks the rotation symmetry, the superfluid stiffness along x and y direction of the lattice are different. Therefore we calculated the stiffness and Drude weight along both the lattice directions in this phase. For a similar striped SS phase, superfluidity has been shown to be lost via a Kosterlitz-Thouless type transition~\cite{finiteT} inspite of the coexisting long range Ising type charge order. It was shown that intersection values of $\rho_s(T^\star)=\f{2T^\star}{\pi}$ for different system sizes follow the logarithmic correction $T^\star=T_{KT}[1+1/(2\ln(L/L_0)]$. This indicates weak coupling between the XY field and the Ising order because for a situation where the two fields are strongly interacting, nature of transition is expected to change~\cite{ssKT}.  From fig.~\ref{ss1_T}, the $KT$ transition temperature for the SF order is estimated to be around $T_{KT}\sim 0.28t$. Thus below this temperature we have a homogeneous mixture of a CDW and a SF; along one lattice direction system behaves like a supersolid (SS) while in the other direction it behaves like a SF.

\begin{figure}[h!]
\begin{center}
\hspace{-1cm}
\includegraphics[width=8.0cm,angle=0]{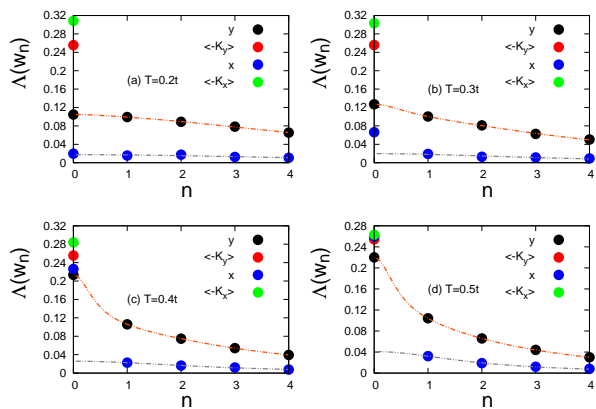}
\caption{Extrapolation plots for $\Lambda(i\omega_n)$ vs $n$ for current along the x and y direction, at various temperatures in the SS-I phase. At low $T$, the extrapolated value of $\Lambda(i\omega_n)$ for both $x$ and $y$ direction, is equal to its value at $\omega_n=0$, and thus $\rho_{x,y}=D_{x,y}$, both being different from the value of the corresponding kinetic energy $\la -K_{x,y}\ra$. For higher temperature values, $lim_{n\rightarrow 0}\Lambda_{xx}(i\omega_n) \ne \Lambda_{xx}(i\omega_n=0)$, which means that $D_x \ne \rho_x$. But along y direction $lim_{n\rightarrow 0}\Lambda_{yy}(i\omega_n) = \Lambda_{yy}(i\omega_n=0)$ and thus $D_y=\rho_y$ even at higher $T$ values.}
\label{ss1_D}
\end{center}
\end{figure}
Left panel in Fig.~\ref{ss1_T} shows the $\rho_x$ and $D_x$ vs $T$ along the $x$ direction for various system sizes and the right panel shows the corresponding data along the y lattice direction. 
Along the x direction, behaviour is very similar to what we saw in the SF phase.  For $T \ll T_{KT}$, $\f{|D_x=\rho_x-\la-K_x\ra|}{\la -K_x\ra} \le 6\%$ while $\f{|D_y= \rho_y-\la -K_y\ra|}{\la -K_y\ra} \sim 48\%$. This implies that the y direction response is far from the linear spin wave approximation, within which $\rho_s\sim D \sim \la -K_x\ra$ (due to vanishingly small contribution from $\Lambda$), even at lowest temperatures. In the model we are studying, at low temperatures this can happen only if the Ising type long range order is getting established.  We already saw in the CDW phase $\rho_s=D$ because $\Lambda$ in both the limits of Eq.\ref{rhos} and \ref{D} is non zero and is equal to $\la -K_x\ra$. 

 As we increase $T$ above $T_{KT}$, $D_x$ shows a very slow decrease with $T$ following $\la -K_x\ra$ and remains non zero even when $\rho_x$ has dropped to zero at $T_{KT}$, in complete analogy with other 2d SF phases. But $D_y$ remains equal to $\rho_y$ even at higher temperatures and goes to zero at $T_{KT}$ along with $\rho_y$. This can be understood in terms of the theorem from Scalapino et. al~\cite{Scalapino} which states that in a system with a spectral gap $\rho_s=D$. Since in the striped SS phase, CDW order exists only along y direction (in the thermodynamic limit), the spectral gap must be anisotropic being non zero only along the y direction. Detailed plots of the current-current correlation function $\Lambda(i\omega_n)$ in support of this observation are shown in Fig.~\ref{ss1_D}. Notice that along $y$ direction $lim_{\omega_n\rightarrow 0}\Lambda_{yy}(i\omega_n) = \Lambda_{yy}(\omega_n=0)$ at all temperatures while $lim_{\omega_n\rightarrow 0}\Lambda_{xx}(i\omega_n)< \Lambda_{xx}(\omega_n=0)$ for $T> T_{KT}$. 
\begin{figure}
\begin{center}
\includegraphics[width=7.5cm,angle=0]{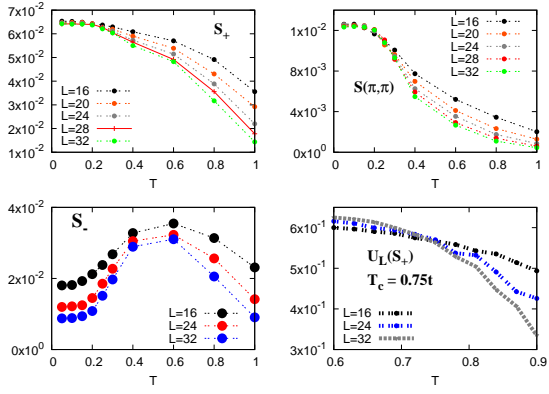}
\caption{Left panels show $S_{\pm}$ vs $T$ respectively while the top right panel shows $S(\pi,\pi)$ vs $T$ for various system sizes. Binder cumulants for $S_{+}$ are shown in last panel which shows a transition temperature of $T_{cdw}\sim0.75t$.}
\label{ss2_S}
\end{center}
\end{figure}
Therefore, in the normal phase of this anisotropic SS-I phase, the system behaves like an ideal conductor along x direction while it behaves like an insulator along y direction.   
\begin{figure}
\begin{center}
\includegraphics[width=7.5cm,angle=0]{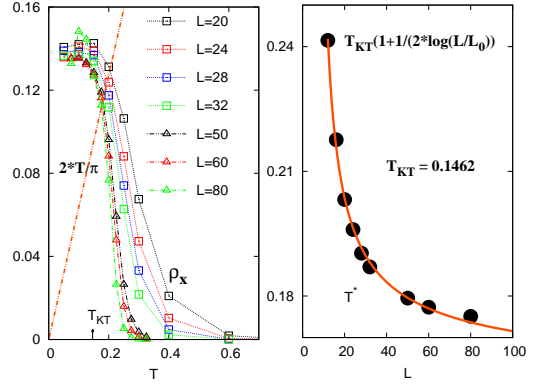}
\caption{Left panel shows $\rho_x$ vs $T$, for the SS-II phase, for various system sizes. The point at which $\rho_x$ vs $T$ curve crosses $2T/\pi$ gives $T^\star$. Right panel shows $T^\star$ vs $L$ is fitted well with the function $T^\star=T_{KT}[1+1/(2\ln(L/L_0))]$ and gives the KT transition temperature to be $T_{KT}=0.146t$.}
\label{ss2_KT}
\end{center}
\end{figure}
\subsection{Supersolid-II}
Another SS phase can be realized in model in Eq.~\ref{hamil} for the set of parameters $t=0.9,t'=0.1,V_1=4.5,V_2=4.5,h=11.5$, as reported in ~\cite{Melko}. This is a quarter empty star SS phase which has a ground state characterized by non zero $S(\pi,\pi)$ and $S_{+}$ shown as a function of $T$ in Fig.~\ref{ss2_S} for various system sizes. There is a weak anisotropy in the CDW order at $T=0$ as clear from very small values of $S_{-}$ compared to that of $S_{+}$ at $T=0$. Interesting feature of the CDW order in this phase is that though $S(\pi,\pi)$ and $S_{+}$ decrease with increase in $T$, anisotropy parameter $S_{-}$ increases with $T$ showing its maximum around $T=0.5t$ as shown in the bottom panel of Fig.~\ref{ss2_S}. To see whether the CDW to normal phase transition is continuous we calculated the 4-th order Binder cumulant $U(S_{+})=1-\frac{<{O^{+}}^4>}{3<{O^{+}}^2>^2}$ where $O^+ = \f{1}{N}\sum_i S_z(i)[\exp(i~\pi~x_i)+\exp(i~\pi~y_i)]$ is the order parameter corresponding to the structure factor $S_+$. Here $(x_i,y_i)$ are corrdinates of site $i$. As shown in the right bottom panel of Fig.~\ref{ss2_S}, data for different system sizes cross each other at $T_c=0.75t$.
\begin{figure}
\begin{center}
\includegraphics[width=8cm,angle=0]{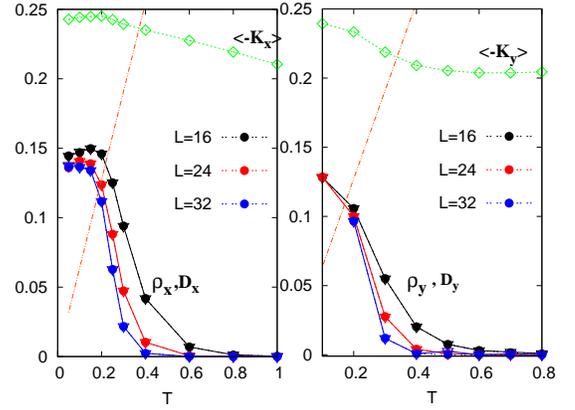}
\caption{SS-II phase: Left panel shows $\rho_x$ and $D_x$ vs $T$ for various system sizes and the right panel shows the corresponding data along the $y$ direction. Note that at all values of $T$, $\rho_{x,y} = D_{x,y}$. Even at low temperature $\rho_{x,y}=D_{x,y} \ne \la -K_{x,y} \ra $ due to the coexisting CDW order in this system.  Both $\rho_{x,y}$ and $D_{x,y}$ drop to zero at $T_{KT}\sim 0.146t$.}
\label{ss2-T}
\end{center}
\end{figure}

After characterizing the CDW order in this phase, we show finite temperature results for the superfluid stiffness in Fig.~\ref{ss2_KT}. Plotting $\rho_{x}$ as a function of $T$, the intersection values of $\rho_{x}(T^\star)=\f{2T^\star}{\pi}$ for different system sizes follow the logarithmic correction $T^\star=T_{KT}[1+1/(2\ln(L/L_0))]$ as shown in right panel of Fig.~\ref{ss2_KT}. We found $T_{KT}=0.146t$ for the superfluid order along $x$ direction.  KT nature of transition again indicates weak coupling between the XY field and the Ising order because for a situation where the two fields are strongly interacting, nature of transition is expected to change~\cite{ssKT}. We expect the same physics to hold for the superfluid order along the y direction. 

\begin{figure}[h!]
\begin{center}
\hspace{-1cm}
\includegraphics[width=8.0cm,angle=0]{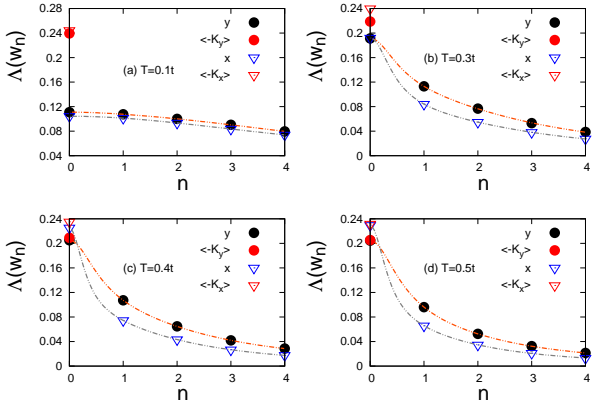}
\caption{Extrapolation plots for $\Lambda(i\omega_n)$ vs $n$ for current along the x and y directions at various temperatures in the SS-II phase. For all values of $T$, the extrapolated value of $\Lambda(i\omega_n)$ for both $x$ and $y$ direction, is equal to its value at $\omega_n=0$, implying that $\rho_{x,y}=D_{x,y}$.  At low $T$ both are different from the value of the corresponding kinetic energy $\la -k_{x,y}\ra$. For $T > T_{KT}$, the extrapolated values of $\Lambda_{xx,yy}(i\omega_n)$ starts approaching $\la -K_{x,y}\ra $. Hence both the stiffness $\rho_{x,y}$ and $D_{x,y}$ are zero for $T > T_{KT}$.}
\label{ss2_D}
\end{center}
\end{figure}

Finally we show the Drude weight calculated along both the lattice directions. 
As shown in Fig.~\ref{ss2-T}, $D_{x,y}=\rho_{x,y} \ne \la -K_{x,y}\ra$ even at very low $T$ due to co-existing CDW order. Detailed plots of the current current correlation function $\Lambda(i\omega_n)$ in support of this are shown in Fig.~\ref{ss2_D}, for various temperatures. With increase in $T$, anisotropy in the CDW order increases and it is reflected in values of $\rho_x$ and $\rho_y$ being different slighlty. Interestingly, along both the directions $D_{x,y}=\rho_{x,y}$ at all temperatures and both the quantities go to zero at $T_{KT}$. This happens because $\lim_{\omega_n\rightarrow 0}\Lambda_{xx,yy}(i\omega_n)=\Lambda_{xx,yy}(\omega_n=0)$ at all temperatures as shown in Fig.~\ref{ss2_D}. Therefore, the normal phase of this SS-II phase is not an ideal conductor.  
\section{Conclusions and Discussions}
We calculated the finite temperature Drude weight for the superfluid and the supersolid phases realized in a system of hard core bosons. Drude weight and the superfluid stiffness can be obtained from different limits of the Kubo formula. Generally in a metal, $\rho_s=0$ while $D\ne 0$ at zero temperature. In an insulator both $\rho_s$ and $D$ are zero while in a superfluid $\rho_s=D\ne 0$ at zero temperature. At any temperature below the transition temperature of the superfluid, $D$ remains non zero resulting in non-dissipative transport and is believed, conventionally, to go to zero for temperatures above the transition temperature. The question we are asking is, in a SF or a SS phase, do these two quantities always remain equal at all temperatures or can they differ from each other? Is it possible to have a dissipatiionless ideal conductor of interacting bosons where $\rho_s$ goes to zero at certain transition temperature but $D$ remains non zero for a range of temperatures above the transition temperature? In the extended XXZ model of Eq.(\ref{hamil}), we calculated the $\rho_s$ and $D$ using SSE. We found that in 2d, in a superfluid phase, at very low temperatures $\rho_s=D=\la -K_x\ra$. As $T$ increases $D$ starts deviating from $\rho_s$. Above $T_{KT}$, $\rho_s$ drops to zero in the thermodynamic limit, but $D$ remains non zero decreasing much slowly with $T$ compared to $\rho_s$. Thus the normal phase of a superfluid in this system is an ideal conductor. We checked this analysis by looking at various SF phases, specially those for which $T_{KT}$ is small so that our calculation of $D$ is reliable. What is the temperature at which $D$ will go to zero can not be determined from our method because extrapolation error becomes large with $T$. 

Although we do not have full microscopic explanation for this surprising observation, we think that it is related to the nature of the Kosterlitz-Thouless transition. Vortex excitations suppress $\rho_s$ making it to drop to zero at $T_{KT}$, but these excitations do not have significant effect on $D$. Therefore, $D$ remains non zero even above $T_{KT}$ and is basically governed by spin waves only. We further studied the quantum XY model in three dimension in order to check this explanation. In 3d quantum XY model, spin waves are the only relevant excitations which make $\rho_s$ to go to zero continuously at $T_c$. From our calculation we saw that $D$ obtained from SSE follows the scaling form $(T_c-T)^{0.187}$ and goes to zero at $T_c$ along with $\rho_s$. Thus  though the normal phase of a 2d SF is an ideal conductor, it is not true for the normal phase of a 3d SF. 

We also studied exotic supersolid phases in 2d where there is a coexistence of the superfluidity and the CDW order. In SS-I phase, there is a striped CDW order which breaks the rotational symmetry. In this case, in the normal phase ($T > T_{KT}$), the system is an ideal conductor along one lattice direction while it is an insulator along the other direction. In the other SS phase we studied, there is a CDW order along both the directions though the anisotropy gets a bit enhanced with increase in $T$. In this case we saw that $D=\rho_s$ along both the directions and both drop to zero at $T_{KT}$ due to the co-existing CDW order. Thus the normal phase of this SS is not an ideal conductor. 

There are deeper questions to be answered in this context, like what makes the Drude weight non zero in the normal phase of a 2d SF? Typically, in a metal, Drude weight vanishes at finite temperature because the delta function part in $\sigma$ gets broadened due to thermal fluctuations. Exceptions are the integrable or near integrable one-dimensional systems~\cite{Heidarian,1dhubbard,integrable} where the Drude weight can remain finite even at finite temperature either due to conserved currents in the system or a part of the current operator has a finite overlap with one of the local conserved quantities, though recently there have been numerical studies on one-dimensional non-integrable systems, where the current operator has overlap with non-trivial quasi-local conserved quantities, showing non-zero Drude weight~\cite{non-integrable}. The system we studied is far from being integrable. Also in a SF phase, $D$ is expected to be non zero at finite temperatures below the transition temperature. In our results we found surprisingly that even above the transition temperature $D$ remains non zero in a 2d system while it goes to zero in a 3d SF. Therefore, it will be interesting to do a vortex dynamic study and understand why vortices can not suppress the Drude weight to zero at $T_{KT}$. It will also be interesting to calculate the Drude weight for other models (e.g. Bose Hubbard model) in two and three dimensions and see if the 2d SF phase has dissipationless transport above $T_{KT}$. These are the questions for future work.     
\section{Acknowledgements}
One of us (A.G) would like to thank P. K. Mohanty for various useful discussions. We are thankful to the anonymous referee whose critical comments helped in improving the quality of this manuscript. 
\section{Appendix A}
In this appendix, we provide details about the fitting of current-current correlation function $\Lambda(i\omega_n)$. In Fig.~\ref{comp}, we have shown four different fits using following functions:
\begin{figure}[h!]
\begin{center}
\hspace{0.3cm}
\includegraphics[width=3.0in,height=6.0cm,angle=0]{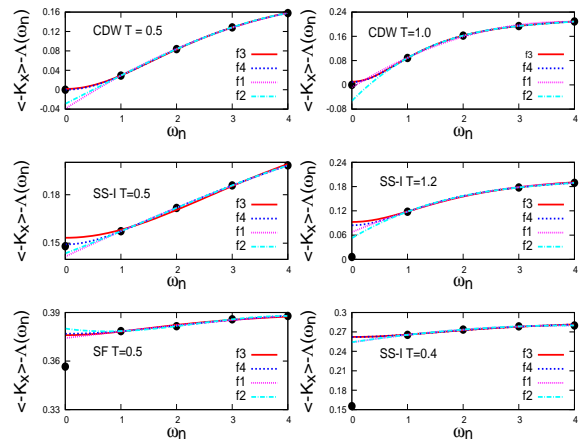}
\caption{$<-K_x>-\Lambda(i\omega_n)$ vs $n$ for the CDW, SS and SF phases for $16 \times 16$ system size obtained from SSE at various temperatures $T$. Top panel shows the results for the CDW phase where $D$ must be zero at any value of temperature. Here Lorentzian provides the best fit to the data and also the physically correct value of $D=0$. Polynomial fit at low $T$ in fact gives negative value of $D$. Similarly in the SF phase shown in the bottom most left panel, polynomial fits give an unphysical uprise in $\la-K_x\ra-\Lambda(i\omega_n)$ near $n=0$ and hence is not acceptable.}
\label{comp}
\end{center}
\end{figure}
\bea
f1(x)=a+b*x+c*x^2 \\
f2(x)=a+b*x+c*x^2+d*x^3 \\
f3(x)=a/(b+x^2)\\
f4(x)=a/(b+x^2)+c/(d+x^2)
\eea
Note that as we know $\la -K_x\ra$, in functions $f1$ and $f2$, $a=\la-K_x\ra$. We first look at the fits for the CDW insulating phase where the answer is known to be $D=0$ at all temperatures. As shown intop panel of Fig.~\ref{comp}, Lorentzian provides the best fit of the data obtained from SSE. At $T=0.5t$, both $f3(x)$ and $f4(x)$ are equally good fits while polynomial functions result in negative value of $D$. At $T= 1.0t$, $f4(x)$ provides the best fit and again polynomial fits result in negative values of Drude weight. 

In the middle panel of Fig.~\ref{comp}, results for the SS phase are shown. At $T=0.5t$, $f4(x)$ provides the best fit. At $T=1.2t$ also $f4(x)$ works well. In the bottom most panel of Fig.~\ref{comp}, results for the SF phase are shown in the left figure. Here the polynomial fit of degree 3 gives unphysical uprise in $\la -K_x\ra -\Lambda(i\omega_n)$ in small $\omega_n$ regime while polynomial of degree 2 gives result very close to what one gets from two Lorentzian fit.

\end{document}